# Identification of Z Cam's Historical Counterpart: The Quest for an Ancient Nova

Bo-Shun Yang[1] and S. M. Hoffmann[1,2]

[1] University of Science and Technology of China, Dept. of the History of Science and Scientific Archaeology, No. 96 Jinzhai Road, 230026, Hefei, China; bsyang@ustc.edu.cn

[2] Friedrich Schiller University of Jena, Faculty of Mathematics and Computer Science, FUSION Group, 07737 Jena, Germany

**Abstract** In recent astronomical discussions, attempts have been made to link the known dwarf nova Z Cam to historical celestial events, particularly the "guest star" phenomenon reported in China in 77 BCE. Despite other suggestions and the problems with regard to the location of the event in 77 BCE, its identification with Z Cam is used in the Variable Star IndeX (VSX) of the AAVSO and in several research papers that aim to derive knowledge on the evolution of cataclysmic variables. Through the reconstruction of the super-constellation of the Purple Palace in the Han Dynasty, we found that Z Cam is actually located outside this enclosure, contradicting the records of the 77 BCE guest star being "within the Purple Palace". With newly found text versions of the guest star in 77 BCE, we narrowed down the position given therein. Combined with a new analysis of accompanying divination text leads to the conclusion that this guest star was actually a comet. Finally, through meticulous examination and comparison, we conclude that the guest star of 369 CE appears the most plausible candidate for Z Cam's historical counterpart, aligning with both textual evidence and modern astronomical observations.

## 1 INTRODUCTION

Z Cam is a known dwarf nova and prototype of the class of Z Cam-type stars. In the recent years, it has attracted considerable attention in the academic community. Shara et al. (2007, 2012, 2024) captured the surrounding shell of Z Cam, demonstrating that it has permitted at least two classical nova eruptions in history. Consistent with the first age estimate of 240 to 2400 years Johansson (2007) linked the most recent eruption of Z Cam to the guest star of 77 BCE. Despite other suggestions (Nickiforov 2010; Warner 2016), astrophysics seems to favour this proposal because Shara et al. (2012, 2024) keep using this identification as the only one. Based on the reading of historical records, Hoffmann (2019) argued that neither Nickiforov's suggestions nor the position of the 77 BCE event match Z Cam's location, and Hoffmann & Vogt (2021) provided a list of some other records that may match the position of Z Cam and had been suggested as potential historical novae by other scholars (although not all these suggestions are convincing). Currently, there are as many as 10 historical records suggested to relate to eruptions of the known dwarf nova Z Cam. With Shara et al. (2024) again referring to the record of 77 BCE which has already been rejected, we now use the most recent results of historical research in order to provide a more profound analysis of this identification and suggest what historical record may preserve observations of the eruptions that caused the shells Shara et al. (2024) have found in the new data.

Additional to the unproven match of Z Cam and 77 BCE in the literature, the textual base of this suggestion may be seen critical: The text version that Johansson (2007) used for the suggestion is Ho (1962) who gives a mediaeval (Song Dynasty, cf. Tab. 1) rewriting of the original observation. Our repositioning of the guest star is based on the original version in combination with newly reconstructed historical star charts with a better epoch (Han Dynasty). Furthermore, the contextual reading of the record provides additional insight in the nature of the object that is described as a guest star.

## 2 DATA ANALYSIS

From the historical point of view, some new findings of the recent years shed new light on both, the record of 77 BCE and the possible identifications of historical records with Z Cam. First, we give our new reconstruction of the constellations that are used as the historical frame of reference; second, we provide new readings of the historical records.

## 2.1 Reconstructing Constellations: the Purple Palace in the Han Dynasty

In all Chinese versions of the recorded observations of the guest star in 77 BCE, it is consistently stated that this guest star was located within the super-constellation of the *Zigong* (紫宫, Purple Palace), that encloses many other smaller constellations. However, the definition of historical constellations have never been canonical but changed over the course of centuries (Hoffmann 2016, 2022; Yang 2023).

All previous astrophysical research utilised a random variety of historical maps. Due to the lack of a comprehensive study of this topic, Hoffmann (2019); Hoffmann et al. (2020); Hoffmann & Vogt (2020) developed their new astrophysical data-mining method with the didactic reduction of a Chinese historical map provided by the Hongkong Space Museum (mainly based upon star charts and maps in the middle Qing Dynasty, 18th century). This version of the Chinese star map placed Z Cam within the right wall of the Purple Palace, seemingly aligning with the description. In contrast, the latest research on the transformations of Chinese constellations (Yang 2023, 297-304) reveals that Purple Palace in 77 BCE (Han Dynasty) has been defined differently and Z Cam is actually located outside the Purple Palace.

During the early Western Han Dynasty (2nd century BCE), the official star system recorded in Sima Qian's *Tianguan shu* (天官书, Book of Heaven Officials) had only 12 stars in the wall of Purple Palace. However, at about 104 to 102 BCE, the Shi school asterism system replaced the old system and the wall of the new Purple Palace contains 15 stars. This new system continued to be transmitted to later generations. Therefore, we can reconstruct the Purple Palace in 77 BCE using star charts or other materials from later periods that are close in time.

Table 1: Chinese Dynasties Used in This Paper.

| Dynasty | years | our reference |
|---|---|---|
| Han | 202 BCE - 220 CE | 77 BCE, 85, 158 CE Figs. 1, 2 |
| Jin | 266 CE - 420 CE | 269, 290, 305, 369 CE |
| Eastern Wei | 534 CE - 550 CE | 537 CE |
| Western Wei | 535 CE - 557 CE | 541 CE |
| Tang | 618 CE - 907 CE | 709 CE |
| Song | 960 CE - 1279 CE | Fig. 1, 2 |
| Jin | 1115 CE - 1234 CE | 1210 CE |
| Qing | 1636 CE - 1912 CE | Fig. 1, 2 |

The star map we use to reconstruct the early Purple Palace is the Cheonsang Yeolcha Bunyajido, which is kept in Korea. This star map was engraved on a stele in 1395, but its base map is a rubbing from a 7th-century star map stele (Stephenson 1976, 560-567). Therefore, the positions and shapes of the constellations preserved on this star map date back to before the 7th century. Pan (2009, 489-495) found this star map to preserve the constellation forms of early Chinese celestial history during the first millennium. Nakamura (2021) calculated the coordinates of the stars on this map and dated them to around 53 BCE, plus or minus 100 years. The constellations depicted on this star map differ significantly in shape and position from those on star maps created after the 11th century but align well with earlier textual descriptions. Recently, we discovered a new star catalogue produced in Han dynasty, and upon comparison, we found it closely matching the Cheonsang Yeolcha Bunyajido.[1] Therefore, this star map is highly reliable to reconstruct the Purple Palace of the Han Dynasty.

Based on this star chart, we reconstructed the Purple Palace and its surrounding asterisms as shown in Fig. 1 (top), which differs significantly from this region in the Song Dynasty (Fig. 1, bottom).

Sun & Kistemaker (1997) have also reconstructed the Han Dynasty sky, but some places in it could be improved. Firstly, the *Shishi xingjing* (石氏星经, Shi School Star Canon) of the Han Dynasty described that "the *Zigong* has 15 stars, among which 7 are on the west (right) wall, 8 are on the east (left) wall (紫微垣十五星，西蕃七，东蕃八)." (Qutan 1980, 666) However, they put only 6 stars at the right wall and 9 at the left. Secondly, this reconstruction does not align with early textual descriptions in the relative position between other asterisms and the wall of the Purple Palace. The original description said "the six stars of *Tianchu* (天厨, Celestial Kitchen) are located at the northeast outside the Purple Palace (天厨六星，在紫微宫东北维外)." (Qutan 1980, 681) But in their star maps, the positions became the east of the Palace wall, similar to the patterns on Song star map (as shown in Fig. 1). The positions of other asterisms, such as the *Wudi neizuo* (五帝内座, Interior Seats of Five Emperors), the *Nüshi* (女史, Female Protocol), and the *Zhushi* (柱史, Official of Royal Archives), also show significant discrepancies with the records in Han dynasty star canons. In contrast, the Cheonsang Yeolcha Bunyajido aligns well with the Han records.

---

[1] This article is accepted and will be published in 2025 on *The Chinese Journal for the History of Science and Technology* (in Chinese).

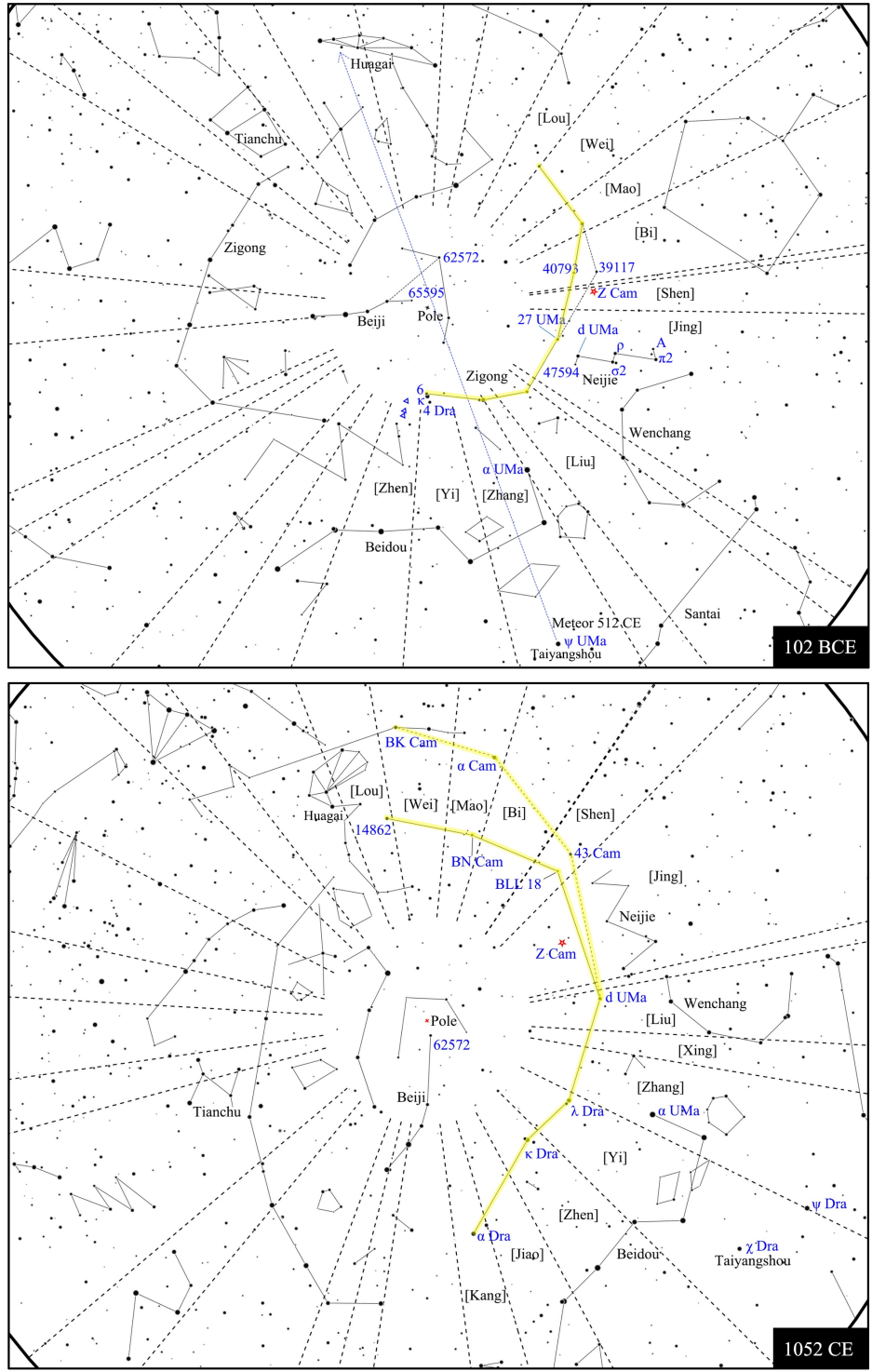

Fig. 1: Purple Palace and Surrounding Asterisms in Han and Song Dynasties (reconstruction by B. Yang according to the originals in Fig. 2). The dashed line the lower map represents the west wall of Qing Dynasty. In the second picture, the mansion boundary line of *Zi* (觜, Beak) is so close to the Shen that they nearly merge into one. Map produced by us.

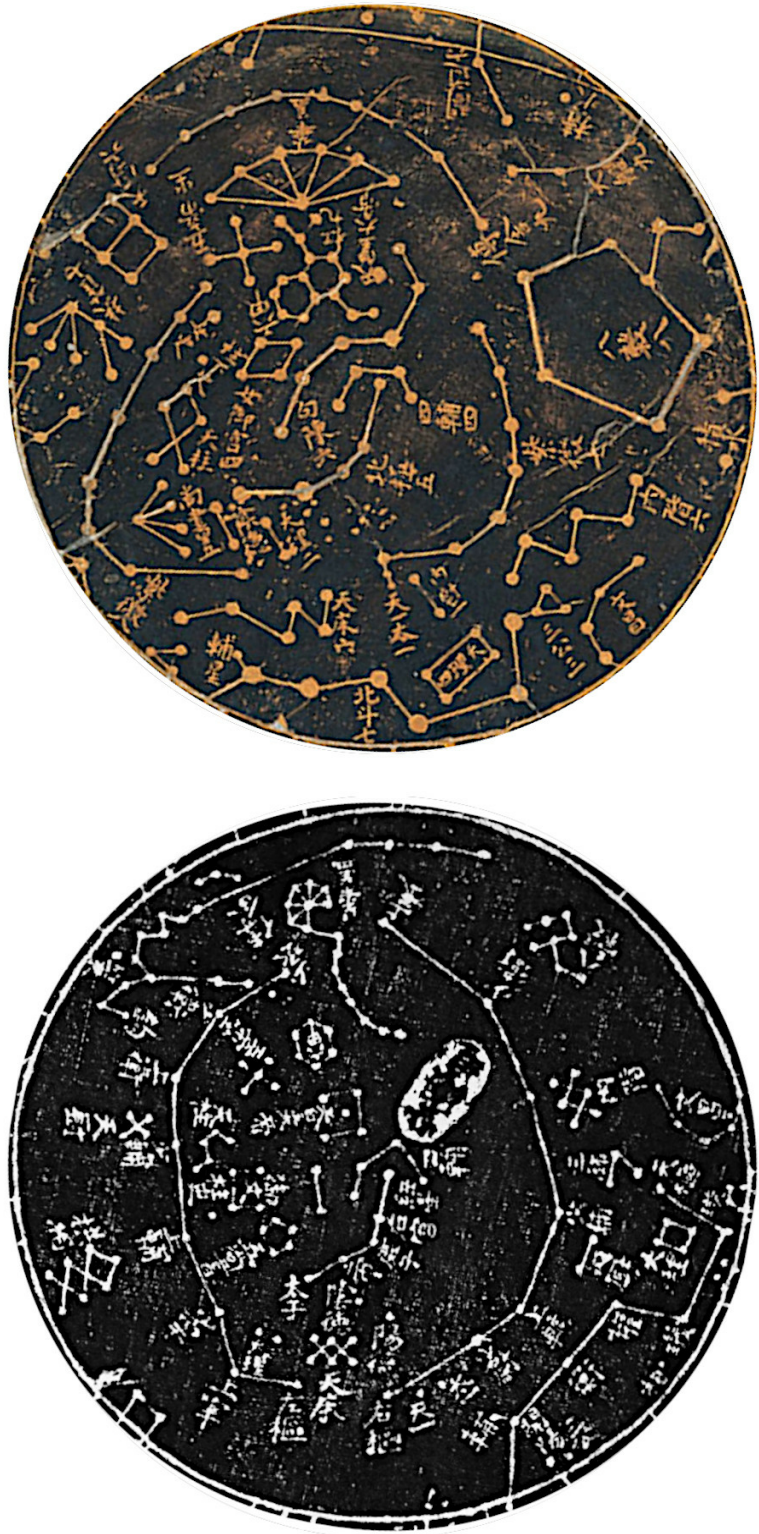

Fig. 2: Purple Palace and Surrounding Asterisms in the *Cheonsang Yeolcha Bunyajido* and *Suzhou Star Chart* (below, produced in Song Dynasty). We chose the same orientation of the Purple Palace with the gate at the bottom and that is how the ancient people set its direction; in the upper map, the Northern Dipper is at the bottom, while in the lower map it is at the right side. Credits: both images produced by us, the upper one derived from a photograph of the woodcut (public domain), the lower one from a screenshot from the zoomable map on https://chinesehsc.org/zoomify/suzhou_planisphere.html (History of Chinese Science and Culture Foundation, Liechtenstein, Jonathan Greet).

Moreover, their reconstruction of the Purple Palace and surrounding constellations seems relied much on Song Dynasty star maps, so the walls of the Purple Palace are much longer compared to the Cheonsang Yeolcha Bunyajido, making the asterism *Huaigai* (华盖, Canopy of the Emperor), which locates at the northern gate of Purple Palace, moves dramatically from its original location. A meteor record from the early 6th century suggests that Sun's reconstruction does not accurately represent the location of *Huagai* in the early Purple Palace area. According to the *Weishu* (魏书, Book of Wei): "In the first year of the Yanchang era, the third month, on the day *yiwei* (April 6, 512), a meteor appeared near the *Taiyangshou* (太阳守, Guard of the Sun), passed through the *Beidou*, entered the Purple Palace, reached the *Beiji*, and extinguished at *Huagai* ((延昌) 元年三月乙未，有流星起太阳守，历北斗，入紫宫，抵北极，至华盖而灭。)." (Wei 1974, 2036) The trajectory of this meteor passed through many asterisms including the *Taiyangshou*, *Beidou*, *Beiji*, and *Huagai*. In Sun and Kistemaker's reconstruction, these asterisms are not aligned in a straight line, which contradicts the record. In contrast, the Purple Palace reconstructed based on Cheonsang Yeolcha Bunyajido, shows that these asterisms are aligned in nearly a straight line, closely matching the meteor record. Additionally, the shape of *Huagai* in this reconstruction also closely resembles the canopy used by the royal family in earlier times.

Based on the clues above, it is evident that the configuration of the Purple Palace depicted in the Korean Cheonsang Yeolcha Bunyajido is generally reliable for the Chinese constellations. However, due to errors introduced during the processes of drawing, transcription, and being engraved on the stone, the star chart is not exact and can only reflect the approximate positions and shapes of the asterisms. Further evidence is required to carefully determine the specific stars corresponding to the walls of the Purple Palace.

Firstly, the most reliable records from the Han dynasty are the star catalogues. During the transition between the 2nd and 1st century BCE, most possibly from 104 to 102 BCE (Ahn 2020; Yang 2023), astronomers of the Western Han Dynasty measured the coordinates of the Shi asterisms, including the Right Star of the Southern Gate of the Purple Palace and its adjacent stars, *Tianyi* (天一, Celestial One) and *Taiyi* (太一, Supreme One), which are indicated with triangles in Fig. 1. The Shi school in Han Dynasty gave a brief description of their relative positions that "*Tianyi* is outside the gate of Purple Palace, south of the right star, at the same degree (in right ascension) (天一星在紫宫门外右星南，与紫宫门右星同度)" and "*Taiyi* is at the south of *Tianyi* and close to it (太一一星，在天一星南相近)", so the three stars should be in the same line and close to each other (Qutan 1980, 667-668). They also mentioned that "When the star *Tianyi* shines brightly and radiantly (which should be its normal state), yin and yang are harmonized, and all things come into being(天一星欲明而有光，则阴阳和，万物成)", from which we know that *Tianyi* should be a relatively bright star compared to *Taiyi*. Based on these data and descriptions, we can identify the three stars respectively as right star of southern gate (6 Dra, m=4.95 mag), *Tianyi* ($\kappa$ Dra, m=3.85 mag) and *Taiyi* (CQ Dra, m=5 mag), which are the same with the identification of Maeyama (2002). The locations and alignment of these stars are consistent with the Cheonsang Yeolcha Bunyajido.

Apart from the right gate star, currently, there is no observational data from the Han Dynasty regarding the other 14 stars of the Purple Palace wall. However, observers from the early 8th century mentioned an old star chart, on which the seven stars of the Purple Palace's right wall were located within the lunar mansions (LM) of *Lou* (娄, Bond), *Mao* (昴, Hairy Head), *Bi* (毕, Net), *Jing* (井, Well), *Liu* (柳, Willow), *Zhang* (张, Extended Net), and *Zhen* (轸, Chariot) (Qutan 1980, 957). The records from ancient star maps inherently contain certain inaccuracies, so the coordinates provided cannot be treated as strictly but should allow for a margin of error. However, it is surprising that five of the seven stars match the positions we reconstructed. The other two stars that don't align precisely are either near the boundary or in an adjacent mansion. This discrepancy is likely due to observational errors. For instance, the right star of the Southern Gate was originally in the *Yi* mansion, but due to errors, Han dynasty observers recorded it as being "ten du into *Zhen* (入轸十度)"[2]. As a result, the star map based on this catalogue also places the star in *Zhen* mansion. The boundaries of these mansions are shown in Fig. 1.

The second and third stars of the right wall of the Purple Palace can be relatively easily identified based on the star maps. However, the fourth star requires careful analysis, as its identification is closely related to the position of Z Cam. Z Cam is located between the fourth and fifth stars from the southern gate of the right wall of the Purple Palace, regardless of whether we consider the version of this constellation from the Han dynasty or the Song/ Qing dynasty. In the Song Dynasty, the fourth star is d UMa (Sun & Kistemaker (1997) chose it as the third star), and the fifth star is HIP 33827 also called BLL 18 (in the Qing Dynasty, it was 43 Cam; Sun & Kistemaker (1997) also chose 43 Cam as the fourth star) which places Z Cam inside the wall. In the Han Dynasty, d UMa is a member of the constellation of *Neijie* (内阶,

Inner Steps), which also differs from its Song/ Qing Dynasty counterpart.

The *Neijie* consists of six stars[3], which form a slanted stair-like shape on Cheonsang Yeolcha Bunyajido and another early star map. In the early 8th century, Tang dynasty astronomers measured this asterism and noted "Now we measured that it is north of *Wenchang*, spanning from the *Jing* to the *Liu* Mansions, narrow and long, its stars close to each other in pairs （three pairs）... (今测在文昌北，东井柳度，又狭长，两相近)." (Qutan 1980, 960) Based on star charts and these descriptions, we can confidently identify the member stars of the Inner Steps: A UMa, $\pi^2$ UMa, $\rho$ UMa, $\sigma^2$ UMa, *d* UMa, and HIP 47594. Due to the precession of the equinoxes, the lunar mansions where these stars are located had shifted by the early 8th century compared to the Han dynasty: the first four stars were in *Jing* Mansion, *d* UMa was in *Gui* Mansion, and HIP 47594 was in *Liu* Mansion, which perfectly aligns with the measurements of Tang dynasty astronomers.

In fact, the *Neijie* was likely modelled after another asterism, *Santai* (三台, Three Steps). In Han dynasty, *Santai* is also called *Tianjie* (天阶, Celestial Steps) or *Taijie* (泰阶, Supreme Steps) among imperial astronomers, described as "*Taiyi* ascending and descending treading on it (太一蹑以上下)". The name "Inner Steps" also implies a staircase, referred to as "the steps of the Heavenly Emperor (天皇之陛也)" (Wei 1974, 532-534). With both constellations resembling stairs, one represents the emperor's passage from *Taiwei* (太微, Supreme Court) to the *Wenchang* (文昌, Administrative Centre) palace, while the other symbolizes the path from the *Wenchang* palace to the Purple Palace.

As a result, the right wall of the Purple Palace during the Han Dynasty can only be between

---

[2] The offset of the triangles from the real position of the identified stars (6 Dra, $\kappa$ Dra and CQ Dra) in Fig. 1 is preserved in the above "old star charts" and in text as "ten degrees into Zhen" (referring to 10°RA, in this case $\sim 2°$ separation); it can be explained by measurement errors with armillary spheres (Yang 2023, 85-108). 365.25 du equals 360°.

[3] Sun & Kistemaker's reconstruction of this asterism contains only 4 stars, which is incorrect.

d UMa and the pole, and the fourth star should be chosen as 27 UMa. No matter whether the fifth star is chosen as HIP 40793 or HIP 39117, Z Cam ends up outside the wall of the Purple Palace in the Han Dynasty, which clearly does not match the record of 77 BCE that reports the guest star "within *Zigong*".

## 2.2 Revisiting the Position of the Guest Star in 77 BCE

Up to now, research on this guest stars is primarily relying on English translations of excerpts of records from Chinese chronicles. There, the original record of 77 BCE reads:

"Emperor Zhao of Han, 4th year of Yuanfeng reign period, 9th month. A guest star was situated within *Zigong* between *Doushu* (斗枢) and *[Bei-]Ji* ([北] 极, North Pole Asterism) (元凤四年九月，客星在紫宫中，斗枢极间)."

This passage was excerpted by Xu et al. (2000, 129) from the Chinese official historical text *Tianwenzhi* (天文志, Astronomical Treatise) in the *Han shu* (汉书, Book of Han). *Zigong* (The Purple Palace) is the circumpolar region, and *Doushu* was regarded to be $\alpha$ UMa (Xu et al. 2000, 422), the first star of the *Beidou* (北斗, Northern Dipper), which is usually called *Tianshu* (天枢, Celestial Pivot). However, this positional description is peculiar: the guest star is located within the Purple Palace, while the reference star "*Doushu*" is outside the Purple Palace. The stars of the western (right) wall of the Purple Palace separate the North Pole Asterism (Beiji) from the reference star $\alpha$ UMa. Why would a star outside the Purple Palace be used as a reference for celestial phenomena inside it?

At the 2023 Chinese Conference of the History of Science and Technology, Zhao Shuaijun proved that there is a textual corruption in the record for 77 BCE because the "Dou" is actually an extra word added by mistake that probably emerged at least in the Song Dynasty, as several printed versions of the *Book of Han* all preserved this error after the Song Dynasty (Zhao 2023). He found that the printed edition of the *Book of Han* after the Song Dynasty differs from the earlier Tang Dynasty manuscript. Therefore, the correct record of 77 BCE needs to be sought in books *before* the Song Dynasty. The *Kaiyuan zhanjing* (开元占经, Divination Canon of Kaiyuan Reign, KYZJ), compiled during the early Kaiyuan reign period (713–741) of the Tang Dynasty, happens to quote the record of the 77 BCE guest star from the *Book of Han* and another book. The quotation from the *Book of Han*, *Astronomical Treatise* reads:

"In the 4th year of Yuanfeng, in the 9th month, the guest star 'qu (去, departed from)' 'Jian' (间, middle or surrounding space) of Shu-Ji (枢极) of *Zigong*. The divination said: 'It indicates war.' In the sixth month of the fifth year, young men were sent from the capital area and other counties/ sub-kingdoms to assist the northern army (元凤四年九月，客星去紫宫中枢极间。占曰：'为兵。'其五年六月，发三辅郡国少年谐北军。)." (Qutan 1980, 781)

Another book called *Hongfan tianwen xingchen bian zhan* (洪范天文星辰变占, The Divination of Star Anomalies in Heavenly Patterns Chapters of Hongfan, HFTWXCBZ), records (Qutan 1980, 781):

"Emperor Zhao of Han, 4th year of Yuanfeng reign period, the 9th month. A guest star 'zai' (在, was in/at) 'Jian' of Shu-Ji of *Zigong*. Three years later, Emperor Zhao died (汉昭元凤四年九月，客星在紫微宫中枢极间。后三年，昭帝崩也)." The two early records are consistent in describing the position of the guest star, without the word "Dou" and by only using "Shu".

The term "Shu" has several meanings. Excluding the first star of the Big Dipper, *Tianshu* (which is outside the Purple Palace), the stars named *Zuoshu* (左枢, left door hinge) and *Youshu* (右枢, right door hinge) in the southern gate of the Purple Palace and the fifth star of the North Pole Aster-ism, *Shuxing* (枢星, pivot star), are known as "Shu" within the Purple Palace; furthermore, the equatorial pole itself could also be called so. The term "Ji" (极, pole) has two meanings: one meaning is the equatorial pole, and the other is the asterism called the North Pole consisting of five stars. The term "Jian" can be interpreted as "middle/ in between" or "surrounding space of something". First, we can exclude *Zuoshu* and *Youshu*, for we find that these names did not appear before the famous astronomical poem *Butian ge* (步天歌, Song of Pacing the Heaven), which was believed to originate from 7th or 8th century (Chen 1992; Pan 2009). Prior to this, the names of the stars of the wall of Purple Palace were different: among them, the left and right star of the southern gate were called *Touguan* (头观) and *Xingui* (信龟) from the Han Dynasty to the late 7th century (Li 1993, 265-268). The fifth star of the North Pole Asterism was named *Shuxing*, which was very close to the equatorial north pole and at that time had the same function as our pole star (Polaris) has today. Thus, many people did not distinguish between the *Shuxing* and the equatorial pole. The pole star in the Song Dynasty was identified to be HIP 62572 (a 5 mag-star between Kochab and Polaris) due to the contemporary star catalogues (Pan 2009, 299), but in 77 BCE, it was about 5 degrees away from the real north pole and cannot be considered the Han pole star. Based on its relatively small separation from the pole in the 1st to the 5th century, Huang (1992) believed that the pole star in that time was HIP 62170 (HD 111112), but at the beginning of Han in 77 BCE, the visible star closest to the true north pole was HIP 65595, and this star is not only brighter than HIP 62170 but also closer to the other four known stars of the North Pole Asterism, so we can regard it as the pole star in 77 BCE, although it is only 5.7 mag bright: it will be recognized near the North Pole Asterism.

The first interpretation of the above records is that this guest star was located between the pole star *Shuxing* (HIP 65595) and the equatorial pole which is an angular distance of ∼ 2° (our: Search Field 1, Fig. 3). Alternatively, we could interpret "Shu" as the equatorial pole, then "Ji" should refer to the five stars of the North Pole Asterism. In this case, the interpretation of the record is that this guest star was located between the North Pole Asterism and the equatorial pole. However, if we consider "Shu-Ji" as a term, which also makes sense, it will refer to the asterism or the pole star or the pole itself. In this reading, "Jian" is no longer "middle/ in between" but refers to the space around a specific object, that is, the space adjacent to the North Pole asterism or the northern pole.

## 2.3 Search for SNRs and nova candidates at the new position

The map in Fig. 3 contains all supernova remnants (SNRs) and pulsars (PSRs) in the area which makes it clear that none of them is within any of our search fields.

The cataclysmic variables (CVs) are only plotted in a circle of 15° around the equatorial pole in 77 BCE. Among them, roughly eight objects in our Search Field(s) have the potential to flare up to naked eye visibility. None of them is in the small search field between the North Pole Asterism *Beiji* and the north pole but the dwarf nova DDE 75 is close to this position. It is of Z Cam-type with dwarf nova peak brightness of 13.8 mag and a quiescence fainter than 18 mag (still, likely too faint). The brightest of all candidates is the dwarf nova SS UMi with a peak brightness of 12.6 mag. In case of a classical nova with typical amplitudes of 11 to 13 mag but possible amplitudes up to 16 mag (Vogt et al. 2019, fig. 1), this object could theoretically flare up to naked eye visibility and reach 4 to 1 mag.

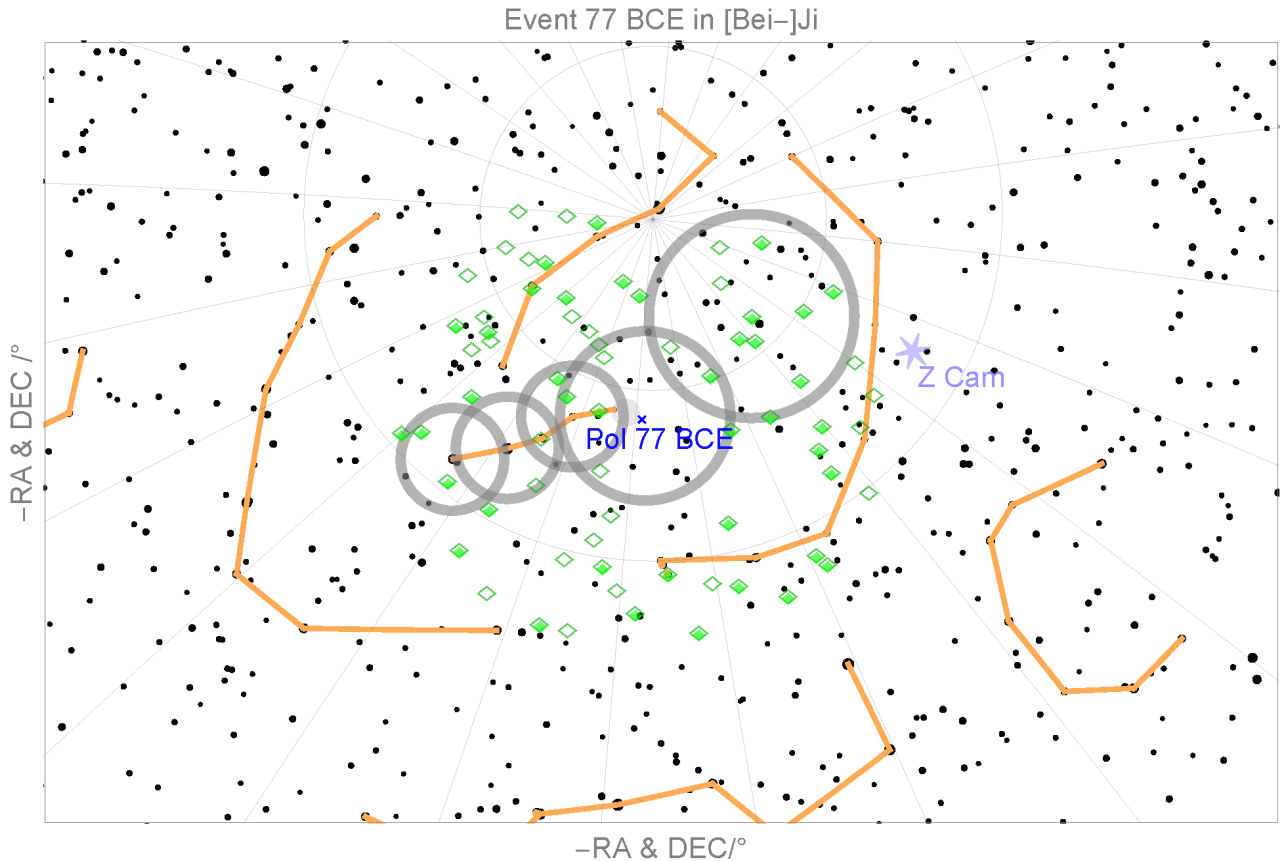

Fig. 3: The Chinese constellations as used in 77 BCE. The circles indicate our search fields in equatorial coordinates, epoch 2000): Search Field 1 is the small gray filled circle at "Pole 77 BCE", Search Field 2 is the area covered by the large (empty) circles whose coordinates can be entered in the VSX-catalogue to find counterpart candidates. The green diamond-symbols indicate all cataclysmic variables around the pole of 77 BCE, those of them brighter than 18 mag in quiescence (found by a filter routine, requiring manual check) are indicated by a filled symbols (our own map, detailed legend in Fig. 6 on page 17).

### 2.4 Is the record 77 BCE really a possible nova?

This record has a long tradition in nova candidate lists. It can be traced to the 19th century and is copied by all authors of the 20th century (Hsi 1957; Ho 1962; Stephenson 1976; Clark & Stephenson 1977; Xu et al. 2000). In the 19th century the term "nova" was used for all kinds of transients that were not understood these days: Stellar evolution has been described from the 1920s onwards, the definition of "supernovae" in astrophysics starts in the 1930s (Trimble 1982, 1983) and "novae" as surface eruptions in cataclysmic or symbiotic binaries were understood as late as the 1960s. Thus, Hsi (1957); Ho (1962) refer to the suggestions of Williams (1871) and Biot (1846) who wrote when it was not even clear what the energy source of stars is, let alone the life cycle of stars or what distinguishes novae from supernovae. Stephenson (1976) copied their suggestion and since then, the record is regarded as a potential (super)nova candidate.

New data from studies of Chinese history challenges this suggestion of the event with regard to the modern definitions of "novae" and "supernovae" in astrophysics.

Among the guest stars mentioned in ancient China, many are comets, and some may even be meteors or aurorae. Xu et al. (2000) attempt to sort out the categories and treat them in different chapters of their two volumes, Stephenson (1976) and his subsequent studies also aim to judge the astrophysical nature by reading records. As they all consider the terminology in chronicles not perfectly reliable, Hoffmann et al. (2020); Hoffmann & Vogt (2020, 2021) recommended the probe the search fields derived from the records for all kinds of variable stars and let astrophysics decide upon the nature of the appearance they saw. Without any supernova remnant in the area, the record would rather be considered a nova than a supernova (or anything else that may or may not be stellar).

However, the inconsistencies of the two records in 77 BCE quoted from the *Astronomical Treatise* and *HFTWXCBZ* in the *KYZJ* are suspicious: The former used the verb "qu (departing)" while the latter used "zai (being in/at)": Are these two records contradictory or is one of them wrong?

The solution requires an understanding of historical writing: both records are correct but the problem is modern extraction of sentences disregarding the context.

The selection criteria for "possible novae" in Chinese records from the 19th century onwards were based on modern standards, aiming to produce lists of potential variable stars and stellar transients of all type, while ignoring the seemingly irrelevant background information. In contrast, ancient Chinese astral science was largely astrology, so abnormal celestial phenomena indicated significant events in the human society. This idea often governed the historiography of early historians: for each recorded celestial phenomenon, they would strive to find the corresponding social event. Although there was no actual connection between the two, ancient historians often established a correspondence between celestial phenomena and human affairs, based on a certain logic such as the formal similarity. Therefore, conversely, we can uncover some information about primitive celestial phenomena from corresponding social events.

The authors of the two books chose different social events corresponding to the celestial phenomena. In the Astronomical Treatise the guest star left the North Pole region, and the government ordered some young men to go to the northern border to assist the army in the following year. We found a record of the same event in the *Annals of Emperor Zhao* of the *Book of Han*:

"In the sixth month, unruly youths from the capital area and other counties/sub-kingdoms, and those who had been officially reported deserters were dispatched to garrison the Liaodong (in northeast Han territory) (六月，发三辅及郡国恶少年、吏有告劾亡者屯辽东)." (Ban 1964, 231)

Therefore, these young men were actually social nuisances, and sending them from the capital to the border corresponds precisely to the guest star leaving the celestial north pole region of the Purple Palace. Therefore, this guest star should be a moving comet rather than a stationary nova.

In the second book, the author chose the death of the emperor three years later as the corresponding event for this phenomenon. Therefore, the author only needed to emphasise that this guest star (a comet) had once been in the northern polar region of the Purple Palace, indicating disaster for the emperor, without mentioning its subsequent whereabouts.

**The record in CE 85 provides a similar case**. The *Astronomical Treatise* in the *Book of Later Han* states:

"2nd year of Yuanhe reign period, the 4th month, day *dingsi* (CE 85 May 20). A guest star appeared in the morning at the east at 8 du within mansion *Wei* (胃, Stomach) measuring three feet in length, and passed by the *Gedao* (阁道, Flying Corridor) into the Purple Palace, staying there for fourty days before extinguishing. The *Gedao* and Purple Palace are the places of the emperor. A guest star that invades and stays for a long time signifies a great mourning. Four years later, Emperor Xiao Zhang died (元和二年四月丁巳，客星晨出东方，在胃八度，长三尺，历阁道入紫宫，留四十日灭。阁道、紫宫，天子之宫也。客星犯入留久为大丧。后四年，孝章帝崩)." (Fan 1965, 3232)

There is no doubt that this was a comet, and because it eventually stayed in the Purple Palace for a long time, it was believed to have caused the emperor's death four years later.

Another chapter in *Book of Later Han, Annals of Emperor Zhang*, also mentioned this comet:

"Emperor Zhang of Han, 2nd year of Yuanhe reign period, in summer, the 4th month, day *yisi*[*dingsi*, May 20][4]. A guest star entered the Purple Palace. On the day of *yimao* (May 18), the emperor's carriage returned to the palace (夏四月乙［丁］巳客星入紫宫。乙卯，车驾还宫)." (Fan 1965, 150)

In this record, the comet is not considered a disaster but symbolises the emperor's carriage, entering the Purple Palace area as the emperor returns from his inspection tour. Here, the historians only recorded the moment of the comet's entry into the Purple Palace and disregarded the movements before and after, as these pieces of information were irrelevant for the correspondence of a celestial event and human affairs they aimed to establish. Through this interesting case, we see the choices ancient historians made regarding the information of primitive celestial phenomena.

With regard to the event in 77 BCE, we can understand why there are records of both "departing" and "being in/at". If the text explicitly mentions movement, it undoubtedly refers to a comet; however, if it only mentions the star being in the Purple Palace, it could be either a moving object or a stationary one. In this case, when the 77 BC comet was associated with the emperor's death, the compiler emphasized its presence in the Purple Palace while omitting any reference to its movement. However, another book, which linked this comet to a different event involving population migration, retained the term "qu (depart from)", providing us with clear evidence that this was indeed a comet.

---

[4] In the 4th month, there is no day *yisi*. This is a corruption and should be corrected to *dingsi*.

**Conclusion 1**: Z Cam does not relate to the guest star of 77 BCE. Actually, even if we consider the wrong text version that dates after the Song Dynasty and places the event of 77 BCE between the $\alpha$ UMa and the North Pole asterism, Z Cam still does not match the position but is 4 or 5 RA-hours away (almost a quarter of the daily revolution (Hoffmann 2019, their fig. 8). The newly found older text version of the record 77 BCE provides a new search field that also does not contain Z Cam.

**Conclusion 2**: The divinatory context of the record additionally suggests that the "guest star" in 77 BCE was rather a comet than a stellar transient.

## 2.5 Searching for the Ancient Nova Corresponding to Z Cam

Since we have now proven that the guest star in 77 BCE is not Z Cam, then which historical guest star could possibly match this known nova? In recent years, scholars have proposed some other candidates for Z Cam (summarized in Hoffmann (2019, tab. 7); Hoffmann & Vogt (2021)). In chronicle order, we can list them below as mostly quoted from Xu et al. (2000):

**CE 85 May 20 [China]** Emperor Zhang of Han, 2nd year of Yuanhe reign period, in summer, the 4th month, day *yisi*[*dingsi*]. A guest star entered the Purple Palace ((汉章帝元和二年) 夏四月乙［丁］巳, 客星入紫宫). [*Hou Hanshu • Zhangdiji*]ch.3

**CE 158 Mar 18 -Apr 15 [Korea]** 13th year of King Chadae of Koguryo, in spring, the 2nd month. There was a Bei (孛, fuzzy) star in *Beidou* ((次大王) 十三年，春二月，星孛于北斗). [*Samguk Sagi*] ch. 15

**CE 269 Oct 13 -Nov 10 [China]** Emperor Wu of Jin, 5th year of Taishi reign period, 9th month. There was a Bei (孛, fuzzy) star in *Zigong* ((晉武帝) 泰始五年九月，有星孛于紫宮). [*Jin shu · Tianwen zhi*] ch. 23

**CE 290 Apr 27 -May 25 [China]** Emperor Wu of Jin, 1st year of Taixi reign period, 4th month. There was a guest star in *Zigong* ((晉武帝) 太熙元年四月，客星在紫宮). [*Jin shu · Tianwen zhi*] ch. 13

**CE 305 Nov 21 [China]** Emperor Hui of Jin, 2nd year of the Yongxing reign period, 10th month, day *dingchou*. There was a Bei (孛, fuzzy) star in *Beidou* ((晉惠帝永興二年) 十月丁丑，有星孛于北斗). [*Jin shu · Tianwen zhi*] ch. 13

**CE 369 Mar 24 -Apr 22 [China]** Duke Haixi of Jin, 4th year of the Taihe reign period, 2nd month. A guest star appeared at the west wall of *Zigong* through the 7th month (Aug 19 to Sep 17), when it disappeared. The prognostication said: "When a guest star guards *Zigong*, it signifies the minister killing the sovereign (海西太和四年二月，客星見紫宮西垣，至七月乃滅。占曰：客星守紫宮，臣殺主)." [*Jin shu · Tianwen zhi*] ch. 13

**CE 537 Jan 27 一Feb 24 [China]** Emperor Xiaojing of Eastern Wei, 4th year of the Yuanxiang [Tianping] reign period, 1st month. A guest star emerged in *Zigong* (至東魏孝靜帝（元象 [天平]）四年正月，客星出于紫宮). [*Wei shu · Tianxiang zhi*] ch. 105

**CE 541 Feb 11 一Mar 12 [China]** Emperor Wen of Western Wei, 7th year of the Datong reign period, 1st month. A guest star was seen at the Purple Palace((西魏文帝大統七年) 正月，客星出于紫宮). [*Xiwei shu · Jixiang kao*] ch. 5

**CE 709 Sep 16 [China]** Emperor Zhongzong of Tang, 3rd year of Jinglong reign period, 8th month, the 8th day *(renchen)*. There was a Bei (孛, fuzzy) star in the *Zigong* ((唐中宗景龍三年八月) 壬辰，有星孛于紫宮). [*Jiutang shu · Tianxiang zhi*] ch. 36

Emperor Zhongzong of Tang, 3rd year of Jinglong reign period, 8th month, 8th day (*renchen*). The emperor dispatched ten envoys to inspect the entire country. There was a Bei (孛, fuzzy) star in *Zigong* ((唐中宗景龍三年八月) 壬辰，遣十使巡察天下，有星孛于紫宮). [*Jiutang shu · Zhongzong ji*] ch. 7

**CE 1210 Feb 26 – Mar 26 [China]** Emperor Weishaowang of Jin, 2nd year of Da'an reign period, 2nd month. A guest star entered the Purple Palace, and its light spread like the shape of a crimson dragon ((金衛紹王大安二年) 二月，客星入紫微中，其光散如赤龍之狀). [*Jin shi · Tianwen zhi*] ch. 20

As discussed in Sect. 2.3, the first guest star (85 CE) is actually a comet and the last entry in the list above (1210 CE) should be an aurora due its shape and colour.

The terms used in the 2nd, 3rd, 5th, and 9th entries (158, 269, 305 and 709 CE) are all described with the Chinese term Bei (孛). Although scholars have previously hesitated to attribute a clear type to this name, it suggests comets rather than novae. It has been argued earlier that this terminology is not always reliable because the term Bei, "fuzzy star" (the most popular translation of Bei), was used in connection with predictions concerning the SN 1572 (Clark & Stephenson 1977; Hoffmann & Vogt 2021). Still, the most likely association with this term is a comet. First, because it has been clearly defined before the 2nd century BCE: "The 17th year of Duke Zhao of Lu (525 BCE), in winter, there was a Bei star near the *Dachen*. What is Bei? It is a comet (冬，有星孛于大辰。孛者何？彗星也)." Gongyang (1999, 506). Second, we conducted an investigation of all ancient Chinese records that mention "Bei"-stars, and calculated the fractions of them that are comets and (super)novae. If any of the following three conditions were met: relative motion compared to stars, presence of a tail, or matching dates with comet records, it was classified as a comet. If it exhibited a clearly stationary state, it was classified as a nova or supernova. We found a total of 131 astronomical events labelled as "Bei", of which 50 had insufficient information for classification (e. g., "A Bei star appeared in the north") and were excluded. Among the remaining 81 records, as Fig. 4 shows, 80 were classified as comets (99 %), with only the one instance in 1572 referring to a supernova.

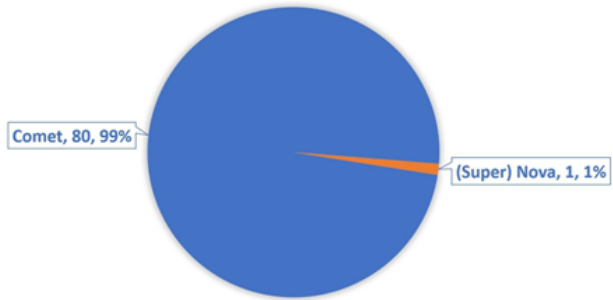

Fig. 4: Statistics on Records of Chinese "Bei"-Stars as Comets or Anything Else.

As the statistical argument alone cannot prove anything for a specific case, we should also mention the finding that before the Song Dynasty, the terms "Bei" and "Hui (broom star)" were essentially used interchangeably. After the Song Dynasty, the "Bei" became less common in astronomical records. In 1572, the supernova was primarily referred to as a guest star, but it was erroneously labelled as "Bei" in reference to its appearance (with rays surrounding it) by using the definition in popular astrological texts of Bei (which is actually "surrounded by many rays (here means many tails) (光芒四出者曰孛)") (Zhu 1425). This might be due to a lack of familiarity among some observers at that time regarding the classification of "Bei". Therefore, at least before the Ming dynasty (1368 CE – 1644 CE), the "Bei" clearly referred to comets.

Additionally, the two records in 158 CE and 305 CE that report "Bei"-stars in the constellation Beidou have already been ruled out by Hoffmann (2019) because the Northern Dipper is outside the Purple Palace and 90°RA away from Z Cam.

The 3rd, 4th, 7th, and 8th entries (269, 290, 537 and 541 CE) are all within the Purple Palace, which does not match Z Cam's location in the new definition of the historical constellation of the Purple Palace at that time. Moreover, *Xiwei shu* (record in 541 CE) is not an official historical record but was compiled by scholars in the 18th century. In the 6th century, the Wei Dynasty was divided into two parts, Eastern Wei and Western Wei. The author of *Wei shu* regarded Eastern Wei as the legitimate authority and thus used the reign titles of Eastern Wei emperors for dating. In contrast, *Xiwei shu* employed the reign titles of Western Wei emperors for dating. The astronomical records in *Xiwei shu, Jixiang kao* were sourced from *Wei shu, Tianxiang zhi*. Given that the 7th year of the Datong reign corresponds to the 4th year of the Yuanxiang reign, it is plausible that the record from 541 CE resulted from the erroneous dating information of the record in 537 CE.

Summarising, among the above ten records, five are comets. Among the remaining five, the three in the constellation of *Zigong* do not match the position of Z Cam (which is outside of and near the western wall of the Purple Palace), and one is highly likely an aurora.

**Conclusion 3**: Only the guest star in CE 369 can be considered as a nova candidate. Apart from the already known information that it lasted for five months, we found the wording "guard" in the following divination text, which indicates that this guest star remains stationary. Its reported position next to the "west wall" is consistent with the position of Z Cam. This event's eruption 1655 years ago is consistent with the observation results presented in Shara et al. (2007) with an age estimate for the shell between 240 and 2400 years, and with an inter-eruption time of more than 1300 years (Shara et al. 2012). Shara et al. (2024)'s new approximate age of the inner shell between 1855 and 4774 yr does not perfectly match the only reported date: the astrophysical dating leads to an eruption between 2749 BCE and 169 CE, and not 369 (which relates to an age of 1655 years), if we take it accurately to a year, but the uncertainties in astrophysics are usually larger. With typical nova amplitudes of 11 to 13 mag, an eruption of Z Cam with a quiescence brightness of roughly 11.8 mag would flare up the star to Sirius' brightness, and it will be visible for the time of decline by $t_3$ to $t_6$. According to Strope et al. (2010), even $t_3$ of roughly 150 days (5 months) are not frequent but occasionally happen, cf. Hoffmann et al. (2020, fig. 2). Decline times by 6 mag of $t_6 \approx$ 5 months would be shorter than currently observed (as $t_6$ typically lasts longer than seven months) but are not totally excluded. As $t_4$ and $t_5$ (which are not used as index) necessarily lay in between $t_3$ and $t_6$, the reported five months in 369 CE do not help to exclude or support the identification with Z Cam. Summarizing, the guest star in CE 369 is the only nova candidate that matches Z Cam in position, although it does not match perfectly in age. If the current astrophysical age estimate again turns out to favour an earlier date, it is well possible that the classical nova eruption of Z Cam in history remained unobserved or that the records are lost as Shara et al. (2007) already stated.

### 2.6 Are there alternative candidates for a nova event in 369 CE?

Hoffmann & Vogt (2020) mentioned that possible eruptions of CQ Dra and BZ Cam may also match the guest star in CE 369. CQ Dra is a symbiotic star which is visible to the naked eye (5 mag) and if it permits a classical nova eruption, this will increase its brightness more than Venus's and it will decline slowly so that a visibility of a couple of months could easily be given. However, from the Han Dynasty to the Tang Dynasty, CQ Dra has always been referred to by astronomers as *Taiyi* (太一, Supreme One). Assuming that they would have noticed that this star had flared up in CE 369, they would not have called it a "guest star" (because a guest is necessarily new and only stays for limited time). So, with regard to the phrasing and framing of the observation in CE 369, an eruption of CQ Dra is unlikely to have been considered as a guest star.

BZ Cam is a nova-like with an impressive multiple shell structure around it. The first photographs from Kitt Peak published in Miszalski et al. (2016) may suggest recurrent nova eruptions on the timescale of centuries or millennia as referred to by Hoffmann & Vogt (2020). However, Balman et al. (2022) with new measurements of the hot flows of this nova-like system (X-ray spectroscopy) suggests that a jet-like outflow mimics an eruption structure that resembles nova shells and their interaction with the ISM even with regard to the ejecta mass. From the historian's point of view, the new studies of the transformations of ancient Chinese constellations that provide better time-resolution suggest that BZ Cam may be considered too far away from the "western wall of the Purple Palace" that is described in the guest star record of 369 CE (see Fig. 5).

The object located directly at the asterism line in Fig. 5 is MGAB-V1226 has been newly discovered in 2020. It passed our brightness filter because it varies around 18 mag (from 17.45 to 18.4), Presuming a hypothetical eruption amplitude of 13 mag or more (up to 16 mag were occasionally observed in some system during the past 250 years, cf. Vogt et al. (2019, fig. 1)), it is not excluded to flare up to naked eye visibility, but it is somehow unlikely for a historical nova (small chance). The new definition of the "west wall" of *Zigong* rather suggests the dwarf nova CI UMa close to this asterism line as a candidate for observations by historical astronomers. This dwarf nova varies normally between 14 and 15 mag (with eclipses to lower brightness) and may therefore become visible to naked eye observers in case of a classical nova eruption with typical amplitudes of 11 to 13 mag; it will likely become roughly as bright as Polaris.

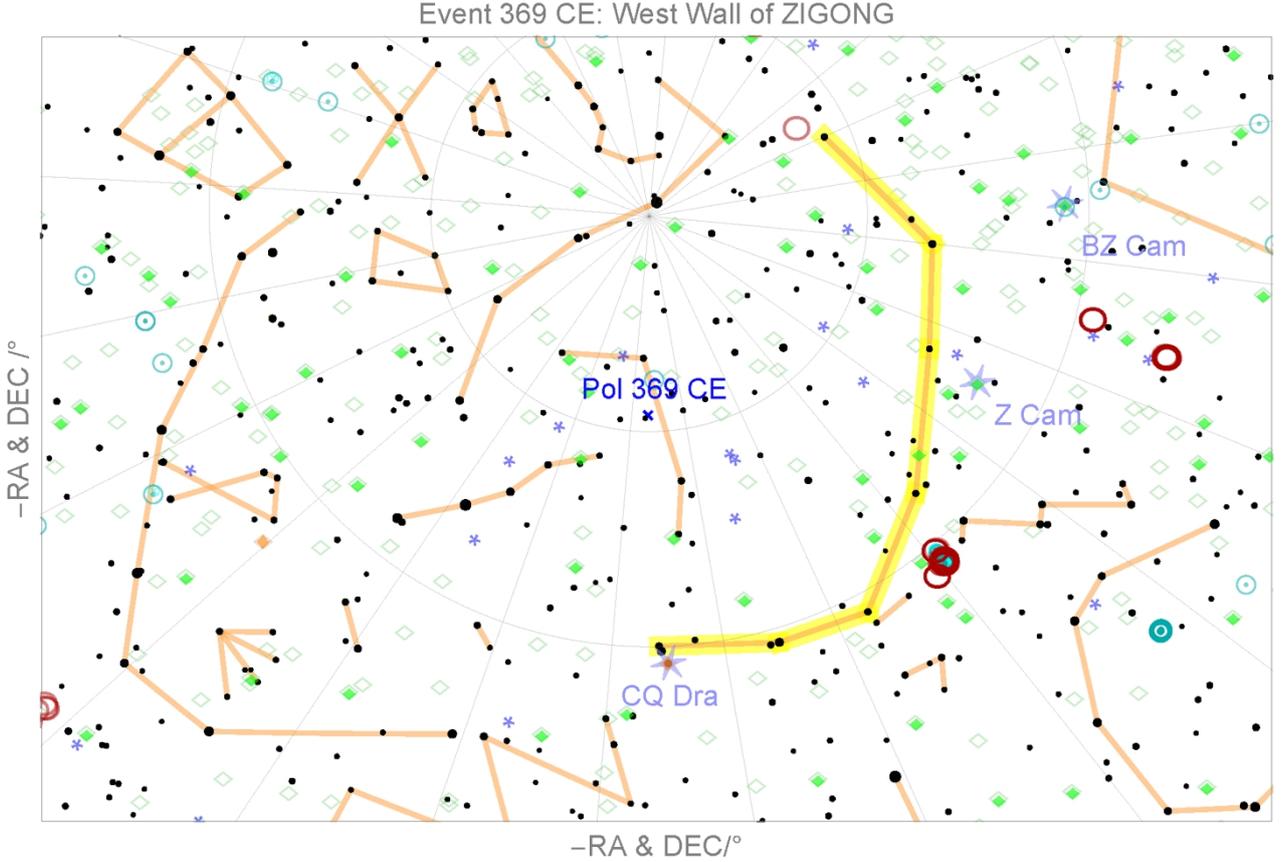

Fig. 5: This map (equatorial coordinates, epoch 2000) places Z Cam with regard to the newly reconstructed Chinese constellations as used in 369 CE. Detailed legend in Fig. 6 on page 17: The green diamond-symbols indicate all cataclysmic variables around the pole of 77 BCE, those of them brighter than 18 mag in quiescence (found by a filter routine, requiring manual check) are indicated by a filled symbols. The red ellipses indicate the known SNRs in the area: There are no SNRs or PSRs young enough in this area to be a supernova counterpart and among the cataclysmic variables, Z Cam is definitely one of the most likely ones (credits: map produced by the authors).

Despite other possibilities that always exist, the inner shell of Z Cam should now be considered the most likely match for this event. The description of the position in this record fits better the historical frame of reference as used in the specific time.

## 3 CONCLUSIONS

As already pointed out by Hoffmann (2019), the position of Z Cam does not match the position of the guest star in 77 BCE. The most recent results of historical research in the comprehensive disser-tation by Yang (2023) concerning the transfer and transformation of Chinese constellations as the historical frame of reference for most "guest stars" yielded a much better time resolution of the definition of them. With this new knowledge, we have to emphasise the rejection of this identification.

Fig. 6: Legends for stars and other symbols in our star charts. "PNc." means "planetary nebula candidate", "CV" means "cataclysmic variable", "bright" indicates the possibility to flare up to naked eye-visibility, "hist.Pos" means "historical position" according to the text.

Furthermore, our profound new reading of the record within its context suggests that the "guest star" in 77 BCE was a comet rather than a stellar transient.

Looking for alternative historical records that could be identified with Z Cam, we support the earlier suggestion by Hoffmann & Vogt (2021) that the "guest star" in 369 which is reported to have lasted for 5 months, could be the counterpart of the innermost shell of Z Cam. This record fits earlier age estimates but is 200 years younger than the youngest possible age in the most recent age estimate by Shara et al. (2024) as long as they do not change their (asymmetric) upper error bar. Thus, it is well possible that the most recent classical nova eruption of Z Cam remained unobserved by historical stargazers or that their record on this is lost – as already suggested by Shara et al. (2007).

# DATA AVAILABILITY

The dissertation by Yang (2023) is currently only available at USTC Hefei and written in Chinese. We are looking forward to a future publication. In addition, we used data from the above-mentioned databases.

**Acknowledgements** We thank our referees for their work. We thankfully used the databases of SIMBAD (Wenger et al. 2000) and ATFN (Manchester et al. 2005), the open source software Stellarium (Zotti et al. 2020), MatLab and Wolfram Mathematica 12.1. We are grateful for the financial aid from the Postdotoral Fellowship Program of CPSF(GZC20232547) and the China Postdoctoral Science Foundation fellowship(2024M753101) for B. Yang, the Otto Neugebauer Fellowship of the Austrian Academy of Science for S. Hoffmann and the project funding by the University of Science and Technology of China (USTC) Hefei. Foremost, we express our warm thanks to Professor Yunli Shi for his support.

# References

Ahn, S.-H. 2020, PASJ, 72, 87 6

Balman, Ş., Schlegel, E. M., & Godon, P. 2022, ApJ, 932, 33 15

Ban, G. 1964, Han shu (The Book of Han) (Beijing: Zhonghua Book Company) 11

Chen, M.-D. 1992, Chenzhuo Xingguan de Lishi Shanbian (The Historical Change of Chenzhuo¡¯s Asterisms), Anthology of History of Science and Technology 1 (Shanghai Science and Technology Press), 77 9

Clark, D. H., & Stephenson, F. R. 1977, The historical supernovae (Oxford, New York, Toronto, Sydney, Paris, Frankfurt: Pergamon Press) 10, 13

Fan, Y. 1965, Houhan Shu (The Book of Later Han) (Beijing: Zhonghua Book Company) 12

Gongyang, S. 1999, Gongyang Chunqiu Zhuan Zhu Shu (Exegesis of the Gongyang Commentary on the Spring and Autumn Annals), Shisan Jing Zhu Shu (Beijing: Beijing University Press) 13

Ho, P. Y. 1962, Vistas in Astronomy, 5, 127 2, 10

Hoffmann, S. M. 2016, Orbis Terrarum, 14, 33 2

Hoffmann, S. M. 2019, MNRAS, 490, 4194 1, 2, 12, 14, 16

Hoffmann, S. M. 2022, in Nuncius Hamburgensis, Vol. 57, Astronomy in Culture - Cultures of Astronomy. Featuring the proceedings of a meeting at the Annual Meeting of the German Astronomical Society (Sept. 2021) (Hamburg/ Berlin: tredition), 491 2

Hoffmann, S. M., & Vogt, N. 2020, MNRAS, 496, 4488 2, 11, 15

Hoffmann, S. M., & Vogt, N. 2021, in Nuncius Hamburgensis, Vol. 55, Applied and Computational Historical Astronomy (tredition), 238 2, 11, 12, 13, 17

Hoffmann, S. M., Vogt, N., & Protte, P. 2020, Astronomische Nachrichten, 341, 79 2, 11, 15

Hsi, T.-T. 1957, Smithsonian Contributions to Astrophysics, 2 10

Huang, Y.-L. 1992, Journal of Tsinghua University (Xinzhu), 15, 93 9

Johansson, G. H. I. 2007, Nature, 448, 251 1, 2

Li, F. 1993, Zhongguo Kexue Jishu Dianji Tong Hui(General Collection of Chinese Science and Technology Classics): Astronomical Parts, Vol. 4, Tianwen Yao Lu (Kaifeng: Henan Educational Press) 9

Maeyama, Y. 2002, in Astrophysics and Space Science Library, Vol. 275, Astrophysics and Space Science Library, ed. S. M. R. Ansari, 3 6

Manchester, R. N., Hobbs, G. B., Teoh, A., & Hobbs, M. 2005, AJ, 129, 1993 17

Miszalski, B., Woudt, P. A., Littlefair, S. P., et al. 2016, MNRAS, 456, 633 15

Nakamura, T. 2021, Journal of Astronomical History and Heritage, 24, 440 3

Nickiforov, M. G. 2010, Bulgarian Astronomical Journal, 13, 116 1

Pan, N. 2009, Zhongguo Hengxing Guance Shi (A History of Chinese Star Observation) (Shanghai: Xuelin Press) 3, 9

Qutan, X.-D. 1980, Yingyin Wenyuange Siku Quanshu, Vol. 807, Kaiyuan Zhanjing (Taibei: The Commercial Press(Taiwan)) 3, 6, 7, 8

Shara, M. M., Mizusawa, T., Zurek, D., et al. 2012, ApJ, 756, 107 1, 15

Shara, M. M., Martin, C. D., Seibert, M., et al. 2007, Nature, 446, 159 1, 15, 17

Shara, M. M., Lanzetta, K. M., Garland, J. T., et al. 2024, MNRAS, 529, 212 1, 2, 15, 17

Stephenson, F. R. 1976, Quarterly Journal Royal Astronomical Society, 17, 121 3, 10

Strope, R. J., Schaefer, B. E., & Henden, A. A. 2010, AJ, 140, 34 15

Sun, X., & Kistemaker, J. 1997, The Chinese Sky during the Han: Constellating Stars and Society 3, 7

Trimble, V. 1982, Reviews of Modern Physics, 54, 1183 10

Trimble, V. 1983, Reviews of Modern Physics, 55, 511 10

Vogt, N., Hoffmann, S. M., & Tappert, C. 2019, Astronomische Nachrichten, 340, 752 9, 15

Warner, B. 2016, in 3rd Annual Conference on High Energy Astrophysics in Southern Africa - Transients, Vol. 241 (Proceedings of Science) 1

Wei, S. 1974, Wei shu (The Book of Wei) (Beijing: Zhonghua Book Company) 6, 7

Wenger, M., Ochsenbein, F., Egret, D., et al. 2000, A&AS, 143, 9 17

Xu, Z., Pankenier, D. W., & Jiang, Y. 2000, East Asian Archaeoastronomy (Amsterdam: Gordon and Breach Science Publishers) 8, 10, 11, 12

Yang, B.-S. 2023, A Research on the Accuracy of Chinese Traditional Star Observation and the Evolution of Constellations, PhD thesis, USTC Hefei 2, 6, 7, 17

Zhao, S.-J. 2023, BC 77 kexing weizhi zaitan (Revisiting the Location of 77 BC Guest Star), Tech. rep., CSHST Beijing 8

Zhu, G.-C. 1425, Tianyuanyuli xiangyi fu (Beijingi: manuscipt of Ming dynasty) 14

Zotti, G., Hoffmann, S. M., Wolf, A., Chéreau, F., & Chéreau, G. 2020, Journal for Skyscape Archaeology, 6(2), 221, https://doi.org/10.1558/jsa.17822 17